%% file: main.tex
\documentclass[
  twocolumn,english,aps,pra,
  superscriptaddress,amsmath,amssymb,floatfix
]{revtex4-2}

\usepackage{amsthm}
\usepackage{amsfonts}
\usepackage{siunitx}
\usepackage{amsmath}
\usepackage{amssymb}
\usepackage{graphicx}
\usepackage{verbatim}
\usepackage[colorlinks]{hyperref}
\usepackage{tikz}
\usepackage{pgfplots}
\usepackage{adjustbox}
\usepackage{braket}
\usepackage{xcolor}
\usepackage{physics}
\usepackage{amssymb} 
\usepackage{graphicx}
\usepackage{dcolumn}
\usepackage{bm}
\usepackage{mathtools}
\usepackage{hyperref}
\usepackage{mathrsfs}
\usepackage{dashrule}
\usepackage{caption}
\usepackage{subcaption}

\newcommand{\be}{\begin{equation}}
\newcommand{\ee}{\end{equation}}

\usepackage[font=small,labelfont=bf,
   justification=justified,
   format=plain]{caption} 
   
\definecolor{linkcolor}{RGB}{0,83,166}
\hypersetup{
  colorlinks = true,
  allcolors = {linkcolor}
}

\begin{document}

\title{Erasing Classical Memory with Quantum Fluctuations: \\ Shannon Information Entropy of Reverse Quantum Annealing}

\author{Elijah Pelofske}
\email[]{epelofske@lanl.gov}
\affiliation{Information Systems \& Modeling, Los Alamos National Laboratory}

\author{Cristiano Nisoli}
\email[]{cristiano@lanl.gov}
\affiliation{Theoretical Division, Quantum \& Condensed Matter Physics, Los Alamos National Laboratory}
\affiliation{Center for Nonlinear Studies, Los Alamos National Laboratory}
\affiliation{Information Science and Technology Institute, Los Alamos National Laboratory}

\begin{abstract}

\input{main_abstract.tex}

\end{abstract}

\maketitle

\input{main_text}

\clearpage

\bibliographystyle{apsrev4-2-titles}
\bibliography{references}

\end{document}

%% file: main_abstract.tex
Quantum annealers can provide non-local optimization by tunneling between states in a process that ideally eliminates memory of the initial configuration. We study the crossover between memory loss and retention due to quantum fluctuations, in a transverse Ising model on odd numbered antiferromagnetic rings of thousands of spins with periodic boundary conditions, by performing reverse quantum annealing experiments on three programmable superconducting flux qubit quantum annealers. After initializing the spins to contain a single domain wall, we then expose it to quantum fluctuations by turning on the transverse Zeeman energy. We characterize the crossover between memory retention at low transverse field, and memory loss at high transverse field by extracting the Shannon information entropy of magnetic domain wall distributions. We demonstrate a clear crossover in memory retention, and its dependence on hardware platform and simulation time. Our approach establishes a general probe of the interplay between quantum fluctuations and memory.

%% file: main_text.tex
Thermal fluctuations can erase memory, and one may ask how and to what extent {\em quantum} fluctuations can induce a loss of information. Understanding how memory of a specific initial state is degraded by quantum fluctuations can illuminate the mechanisms driving quantum phase transitions~\cite{sachdev1999quantum}. Moreover, it is essential for the development of robust and long-lived quantum memory devices~\cite{brown2016quantum,zhao2009long}. Importantly, it advances our understanding of memory effects in quantum annealers~\cite{johnson2011quantum, PRXQuantum.2.030317, perdomo2010studyheuristicguessesadiabatic}. 

The quantum transverse Ising model, introduced by P.~G.~de Gennes~\cite{DEGENNES1963132} over sixty years ago to study  hydrogen-bonded ferroelectrics, has become paradigmatic for quantum phase transitions~\cite{chakrabarti1996quantum}, much like the classical Ising counterpart had been for classical criticality, and has also served as the emergent Hamiltonian for programmable quantum annealers~\cite{Morita_2008, farhi2000quantumcomputationadiabaticevolution, Santoro_2002, Kadowaki_1998}. It is an ideal candidate for our study, because it can be loosely regarded as a classical model perturbed by controllable quantum fluctuations. Its Hamiltonian is composed of two non-commuting Pauli terms,
\begin{equation}
    \mathcal{H} = 
     \sum_{ ij } J_{ij} \hat{\sigma}_z^{i} \hat{\sigma}_z^{j} - \Gamma \sum_i \hat{\sigma}_{x}^{i}.
    \label{equation:Transverse_Ising}
\end{equation}
The first term is diagonal in the $ {\cal Z}$ basis while the second term introduces a transverse Zeeman energy in the ${\cal X}$ basis.  It can therefore be interpreted intuitively as ``classical'' spins aligned along the $ {\cal Z}$ axis, but subject to quantum, rather than thermal, fluctuations controlled by $\Gamma/J$, which serves as a measure of the system's ``quantumness.''

Here we map for the first time the crossover between memory retention  and memory loss as a function of quantum fluctuations. The importance of this regime is highlighted also by recent protocols for performing hysteresis measures~\cite{pelofske2025magnetic} under quantum fluctuations in much more complex and completely different systems; those experiments have shown memory retention for values $10^{-6}<\Gamma/J<10$, similar to our results. We concentrate on the simplest model, the 1D quantum chain on a ring (and thus $J_{ij}=J\delta_{i,i+1}$) as a natural test case that avoids degeneracy lifting due to the transverse field, which can arise in more complex system. It allows us to track the unique topological feature of 1D magnetic domain walls, which has previously been a proposal for quantum annealer dynamics measurement~\cite{king2022coherent, king2023quantum}. We experiment on three different D-wave quantum annealers, in a strongly interacting and almost non-interacting regime. We choose uniformly weighted antiferromagnetic rings of an {\em odd} number of up to $4885$ spins, which necessarily always contain at least one pinned domain wall. We prepare it in an eigenstate of the ${\cal Z}$ basis of one domain wall, and expose it to quantum fluctuations for 1, 100 and 2000 $\mu$s at different values of $\Gamma/J$, and extract the magnetic domain wall Shannon information entropy of the process to quantify memory loss~\cite{shannon1948mathematical}, which is closely related to the thermodynamic properties of quantum annealers~\cite{buffoni2020thermodynamics}.

\begin{figure*}[t!]
    \centering
    \includegraphics[width=0.9\linewidth]{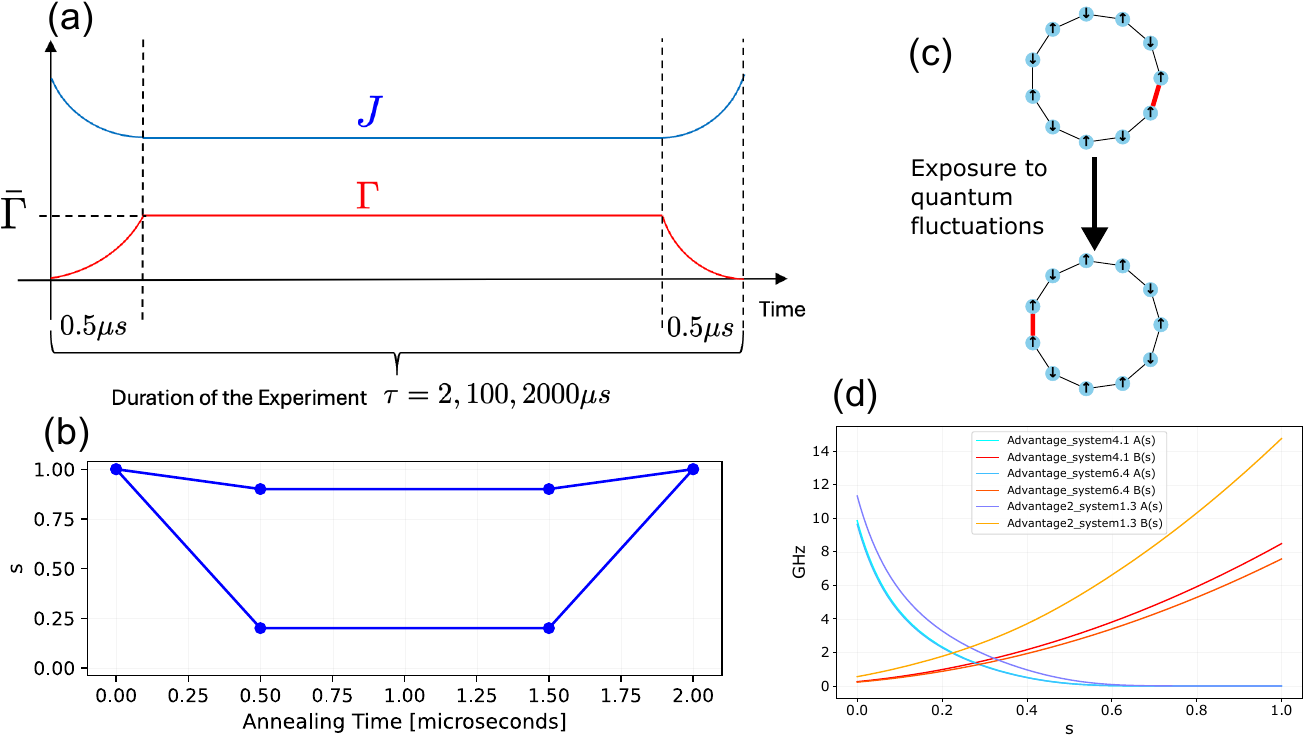}
    \caption{{\bf Schematics of the Experiment.} Panel a: we expose the system to quantum fluctuations by ramping up the transverse field and reducing the coupling, which is achieved by reducing the annealing parameter $s$ and holding it constant, illustrated using two different pause values ($s=0.9, s=0.2$) and an annealing time of $2$ microseconds in panel b. Panel c: The geometrically frustrated pinned domain wall moves after exposure to quantum fluctuations. More domain walls can also form. Panel d: the hardware defined energy scales $A(s)$ and $B(s)$ for the three D-Wave devices. From these we can extract $\Gamma/J=A/(B\bar J)$.  }
    \label{fig:protocol_and_hardware_params}
\end{figure*}

\begin{figure*}[ht!]
    \centering
    \includegraphics[width=0.49\linewidth]{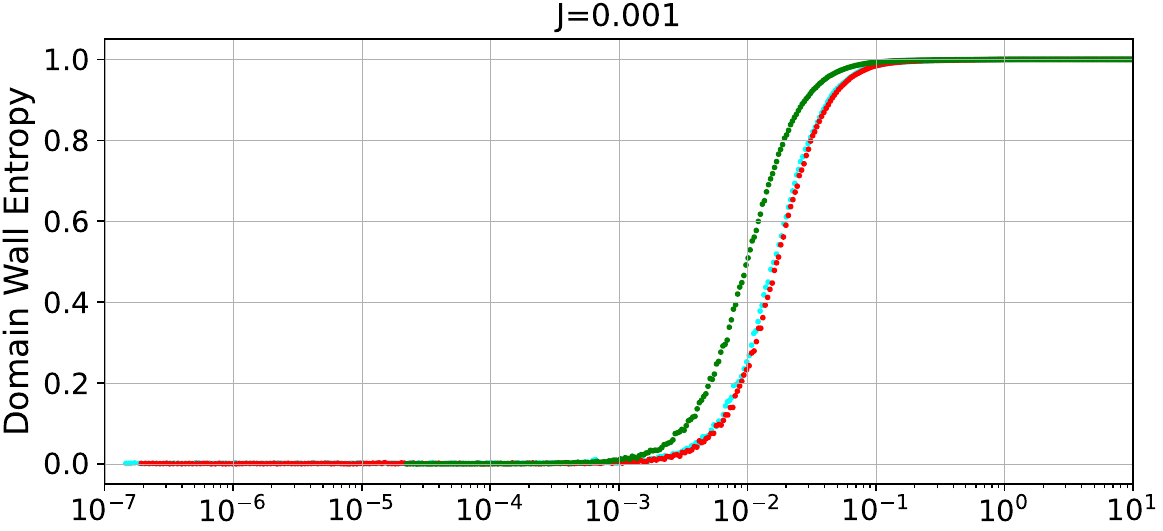}
    \includegraphics[width=0.49\linewidth]{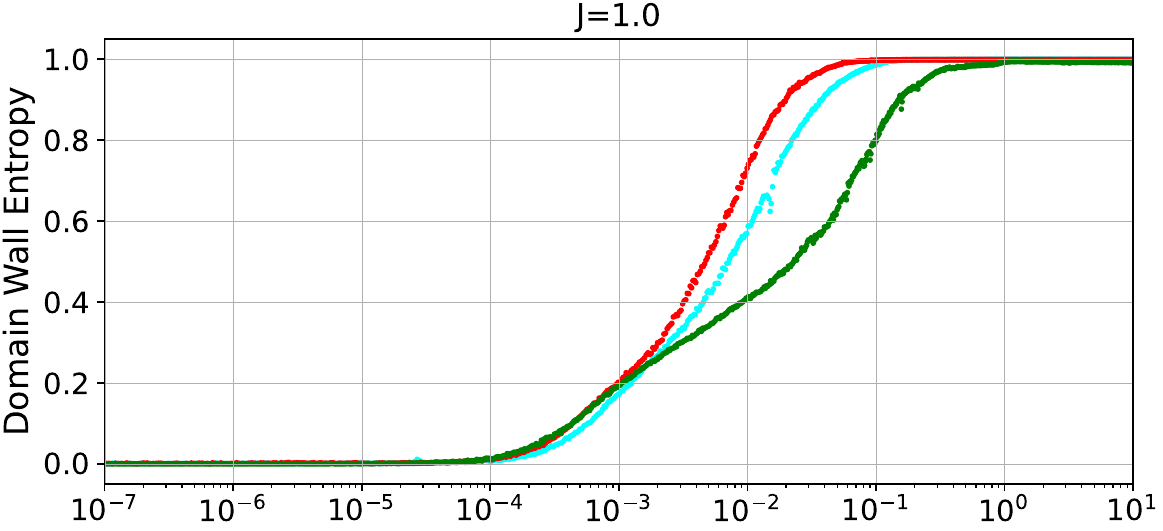}
    \includegraphics[width=0.49\linewidth]{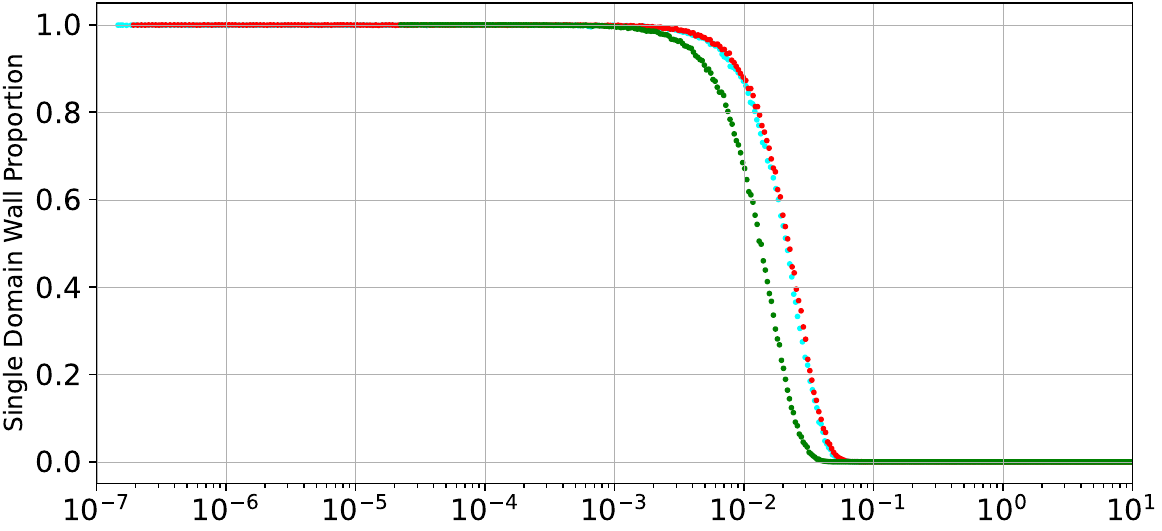}
    \includegraphics[width=0.49\linewidth]{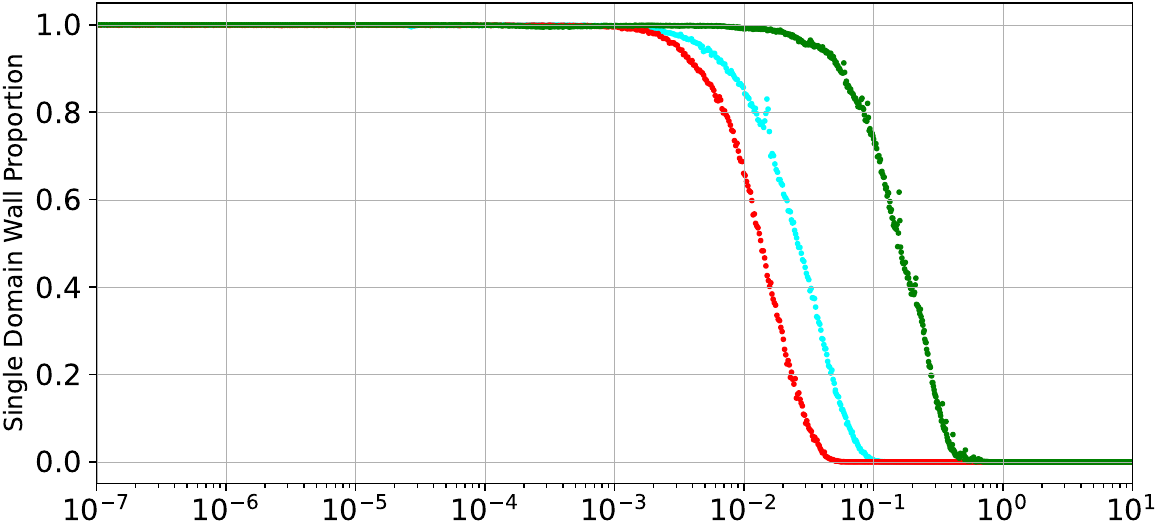}
    \includegraphics[width=0.49\linewidth]{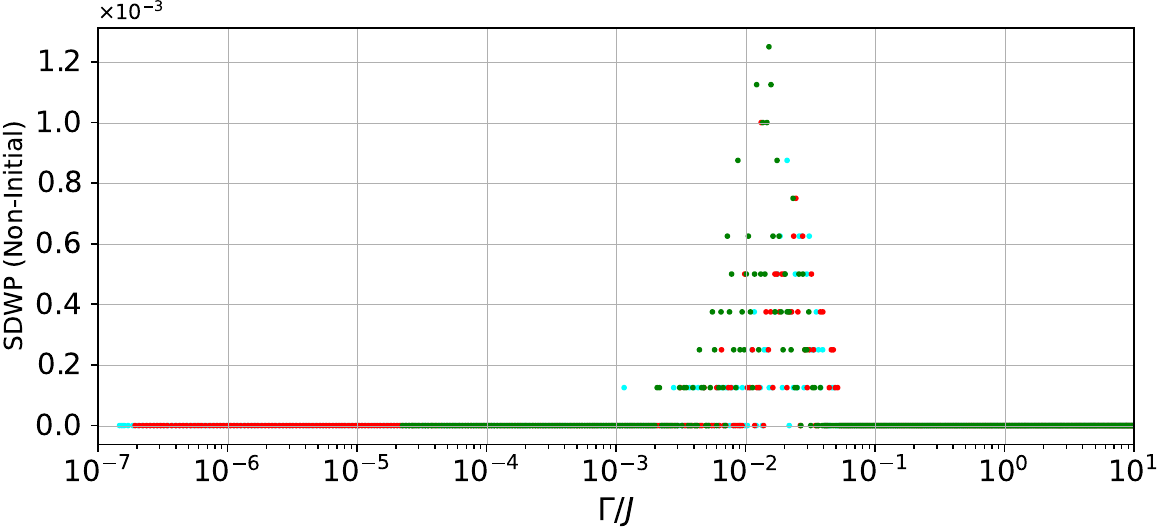}
    \includegraphics[width=0.49\linewidth]{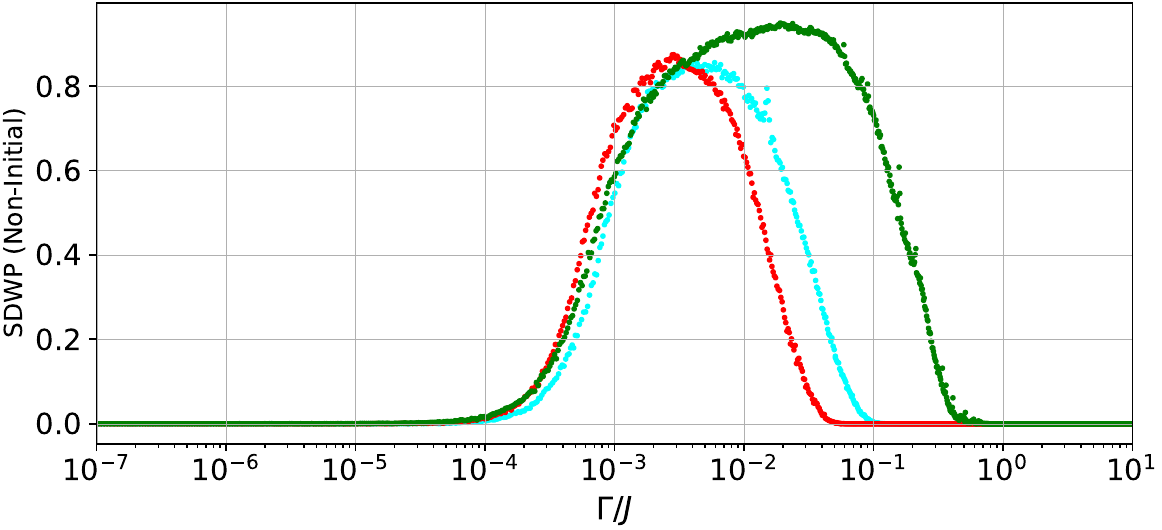}
    \includegraphics[width=0.999\linewidth]{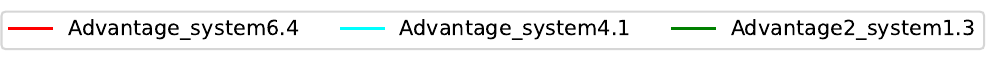}
    \caption{Magnetic Domain Wall Shannon Entropy as a function of the D-Wave QPU anneal-fraction $s$ (plotted as log-scale $\Gamma/J$ on the x-axis), with $2$ microsecond simulation time. Left column shows the weakly coupled $J=0.001$ case, and the right column shows the strongly coupled $J=1.0$ case. The full span of domain wall entropy data is not plotted for visual clarity: we have data down to $
    \Gamma/J \approx 10^{-10}$, where the entropy remains at $0$, and up to $\Gamma/J \approx 10^{4}$, where the entropy remains at $1$. 
    Middle: Total single-domain wall sample proportion (either both spin up $\uparrow \uparrow$ or both spin down $\downarrow \downarrow$). 
    Bottom: Total single-domain wall proportion (abbreviated as SDWP) which moved from the initial pinned position specified by the reverse anneal initial spin configuration (or has the opposite domain wall orientation to that initial domain wall). Notice that in the weakly coupled $J=0.001$ case, the non-initial single domain wall proportion (lower-left sub-plot) is on the order of at most $\approx 1\times 10^{-3}$. }
    \label{fig:domain_wall_entropy_2_ms}
\end{figure*}

\begin{figure*}[ht!]
    \centering
    \includegraphics[width=0.49\linewidth]{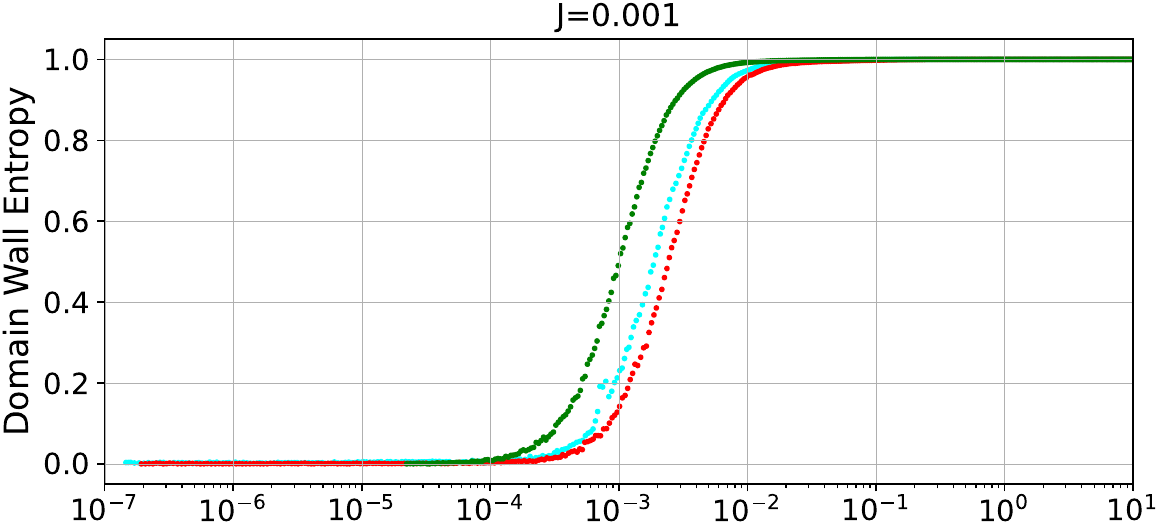}
    \includegraphics[width=0.49\linewidth]{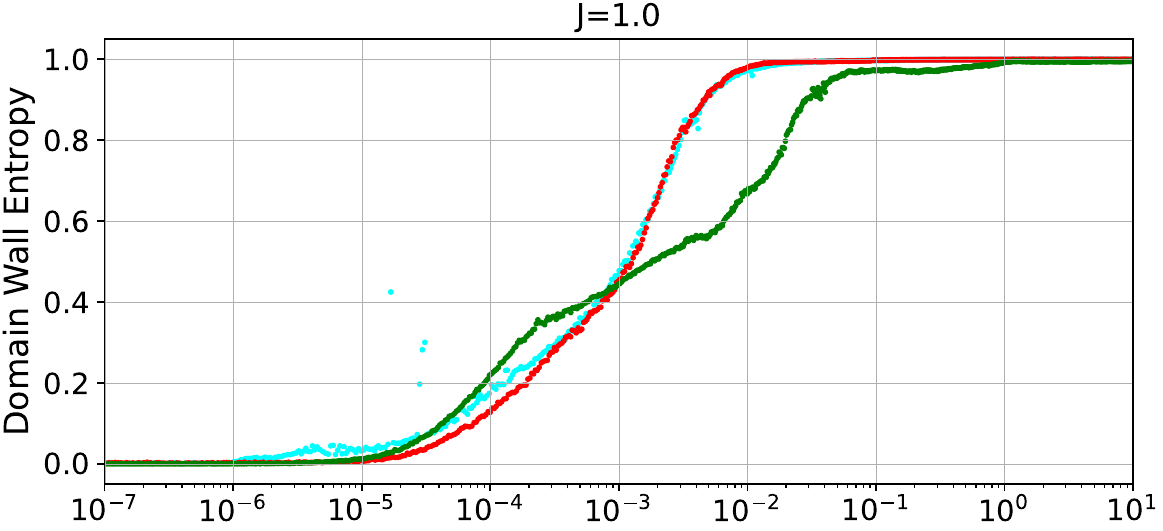}
    \includegraphics[width=0.49\linewidth]{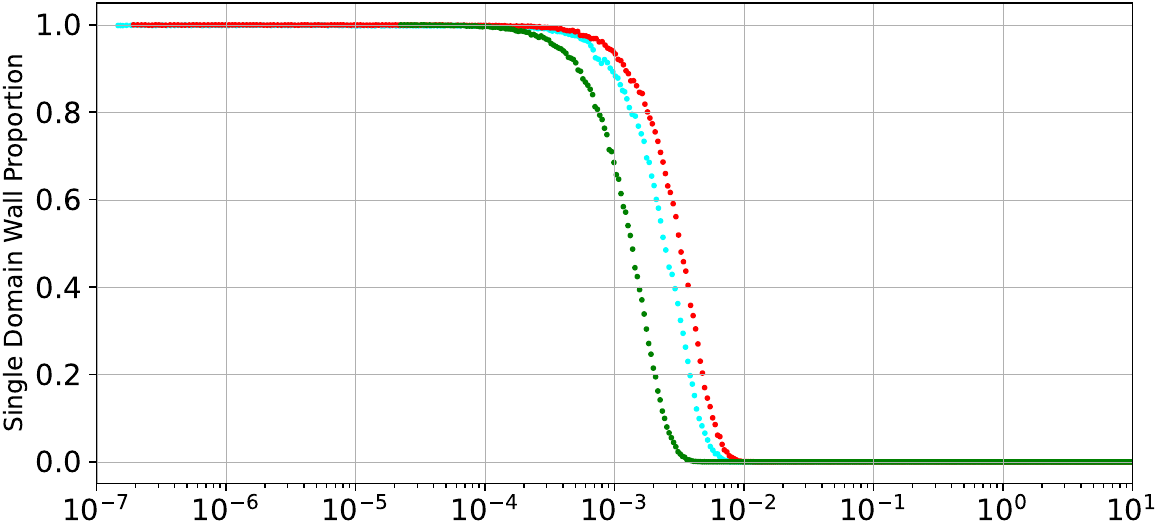}
    \includegraphics[width=0.49\linewidth]{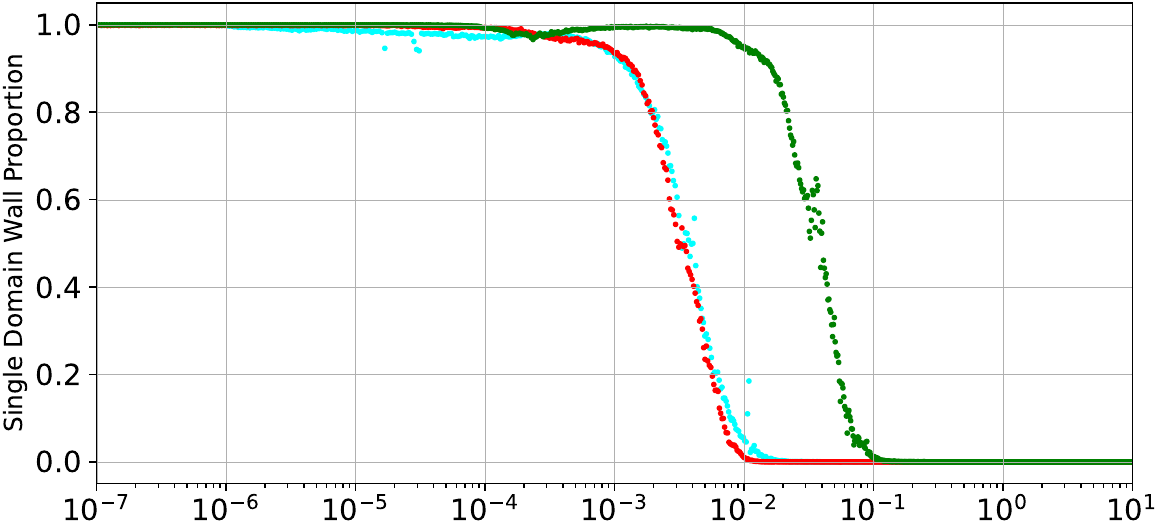}
    \includegraphics[width=0.49\linewidth]{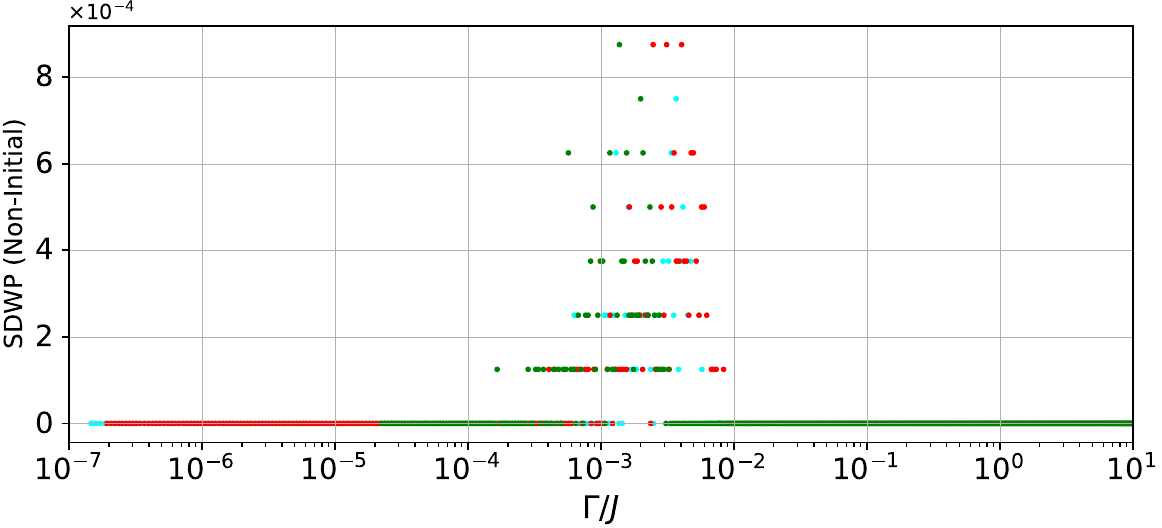}
    \includegraphics[width=0.49\linewidth]{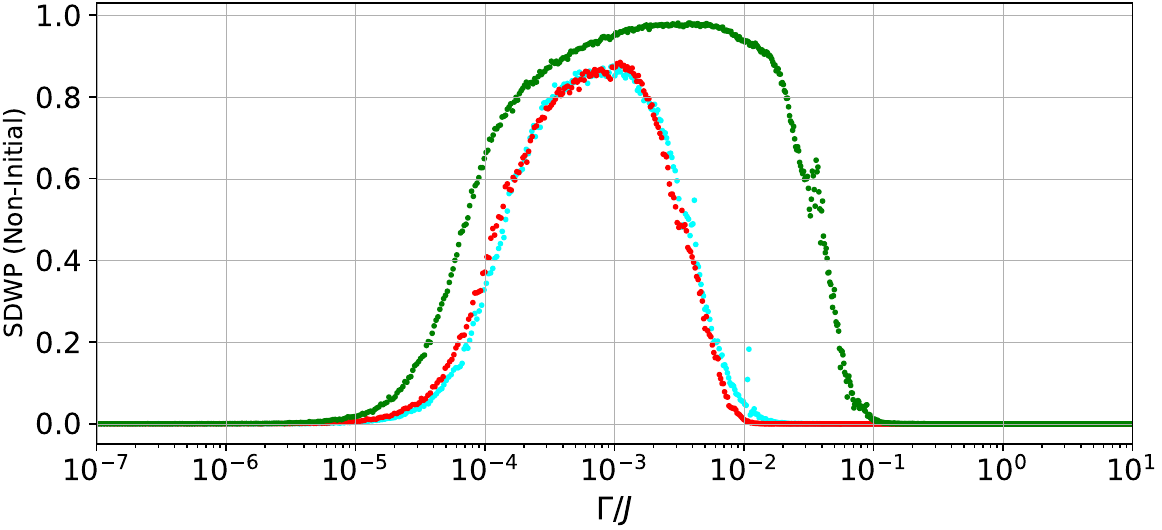}
    \includegraphics[width=0.999\linewidth]{figures/domain_wall_entropy/legend.pdf}
    \caption{Magnetic Domain Wall Shannon Entropy as a function of the D-Wave QPU anneal-fraction $s$ (plotted as log-scale $\Gamma/J$ on the x-axis), with $100$ $\mu$s annealing time. Left column shows the weakly coupled $J=0.001$ case, and the right column shows the strongly coupled $J=1.0$ case. The full span of domain wall entropy data is not plotted for visual clarity: we have data down to $
    \Gamma/J \approx 10^{-10}$, where the entropy remains at $0$, and up to $\Gamma/J \approx 10^{4}$, where the entropy remains at $1$. 
    Middle: Total single-domain wall sample proportion (agnostic to the spin orientation of the single domain wall). 
    Bottom: Total single-domain wall proportion (abbreviated as SDWP) which moved from the initial pinned position specified by the reverse anneal initial spin configuration (or has the opposite domain wall orientation to that initial domain wall). Each sub-plot shares the same x-axis scale.  }
    \label{fig:domain_wall_entropy_100_ms}
\end{figure*}

\begin{figure*}[ht!]
    \centering
    \includegraphics[width=0.49\linewidth]{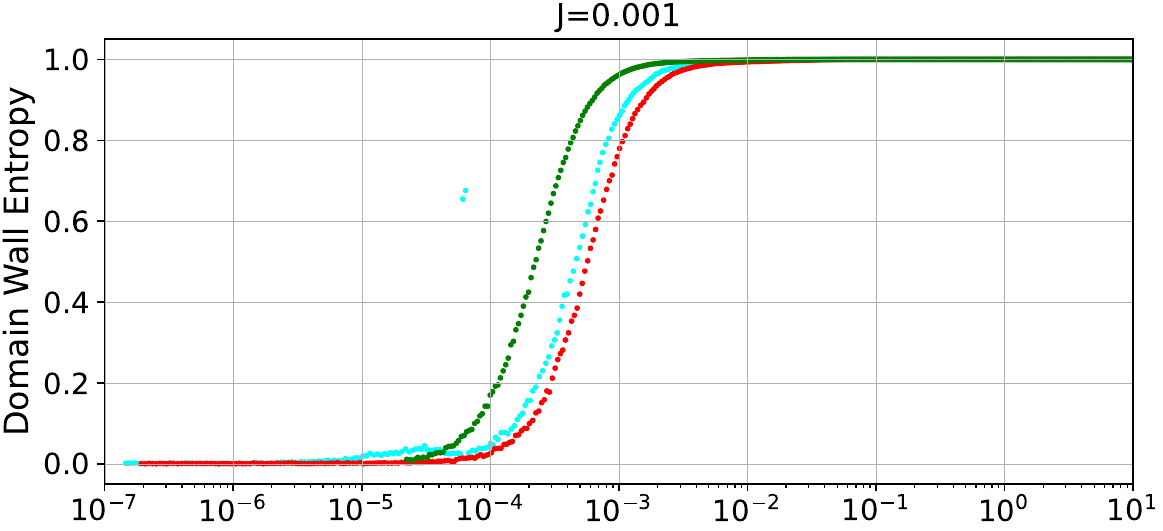}
    \includegraphics[width=0.49\linewidth]{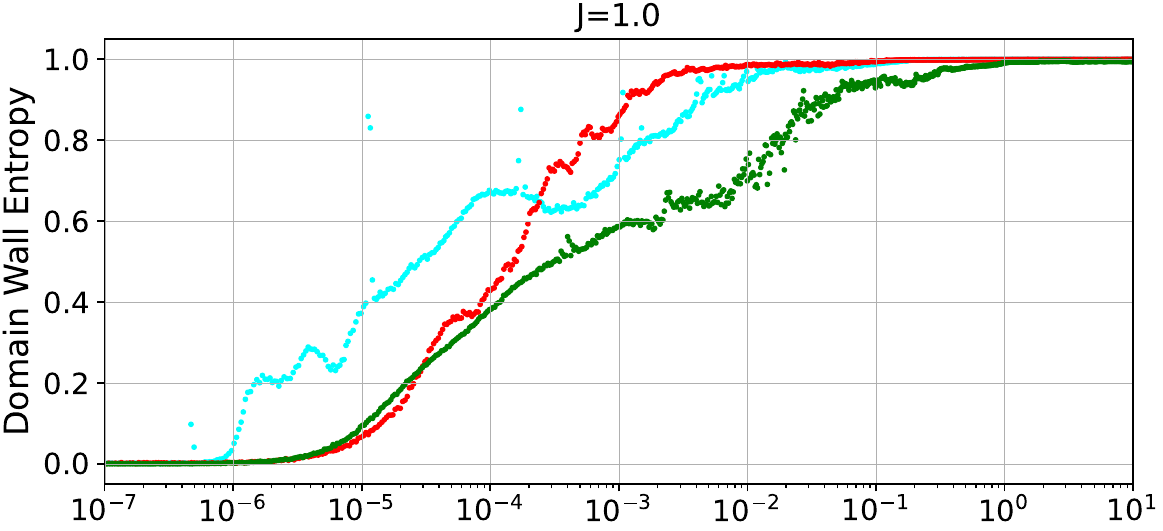}
    \includegraphics[width=0.49\linewidth]{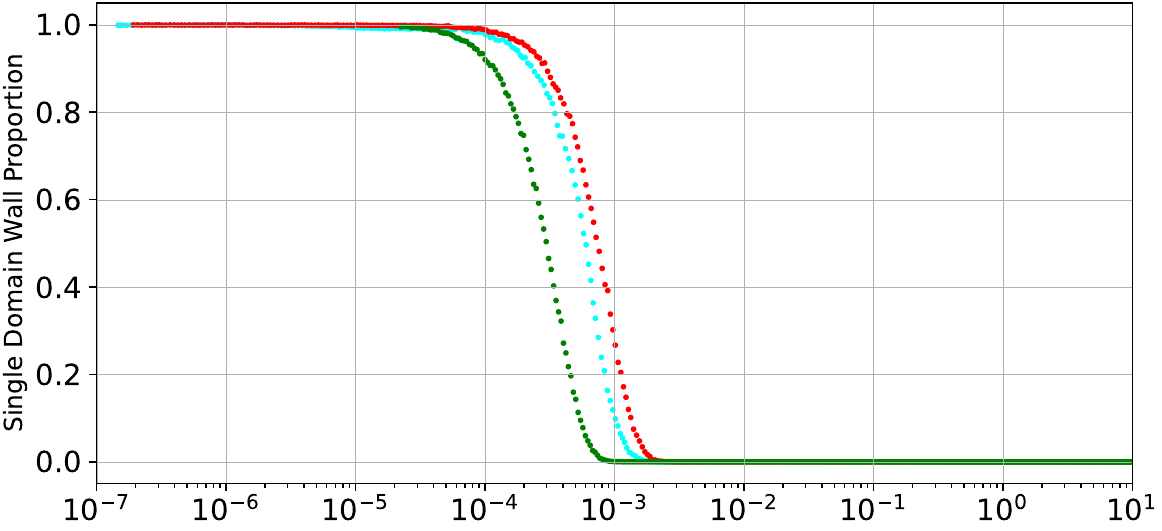}
    \includegraphics[width=0.49\linewidth]{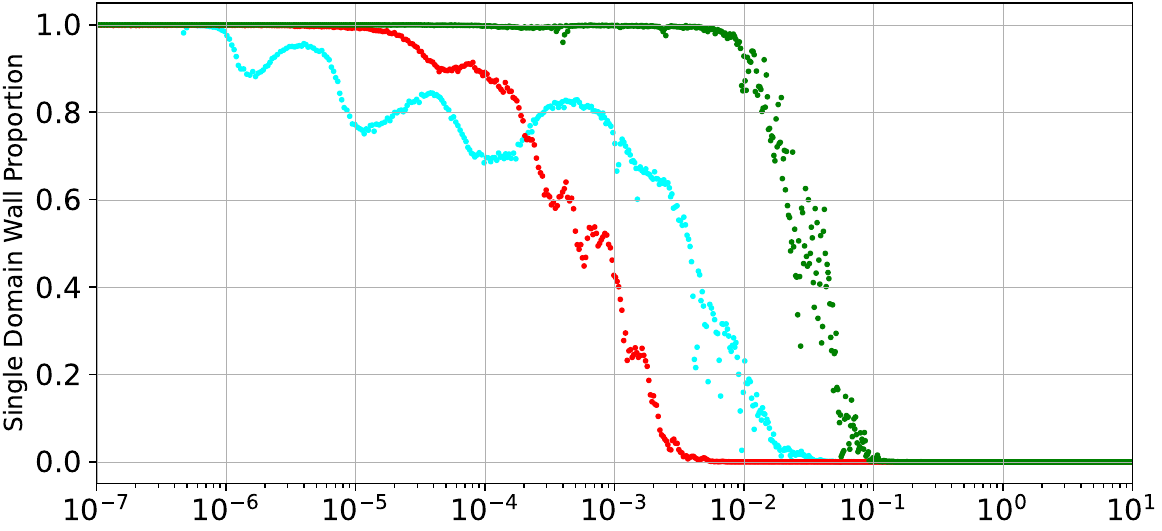}
    \includegraphics[width=0.49\linewidth]{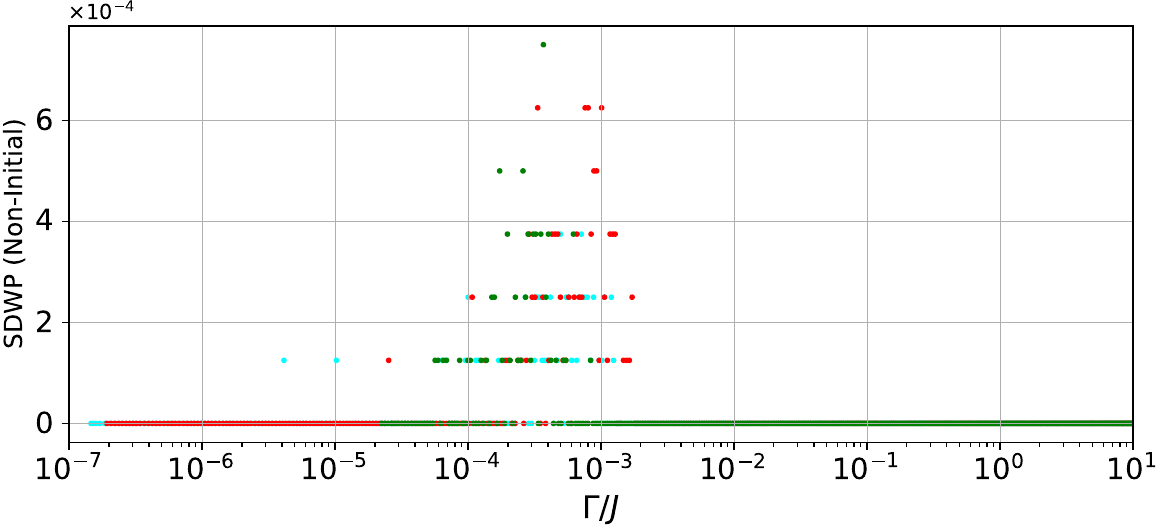}
    \includegraphics[width=0.49\linewidth]{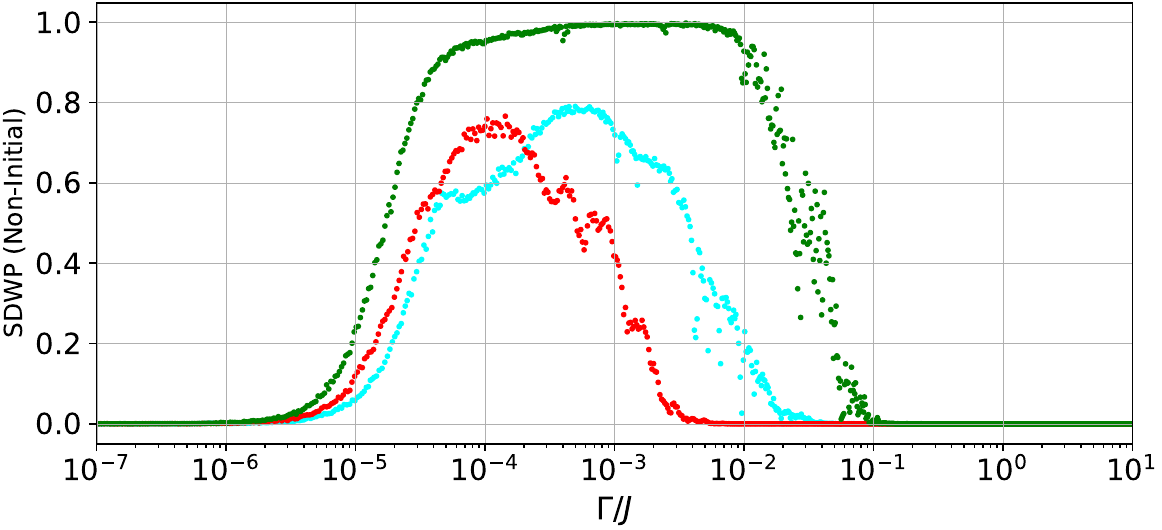}
    \includegraphics[width=0.999\linewidth]{figures/domain_wall_entropy/legend.pdf}
    \caption{Magnetic Domain Wall Shannon Entropy as a function of $s$ (plotted as log-scale $\Gamma/J$ on the x-axis), with $2000 
    \mu$s annealing time. The full span of domain wall entropy data is not plotted for visual clarity: we have data down to $
    \Gamma/J \approx 10^{-10}$, where the entropy remains at $0$, and up to $\Gamma/J \approx 10^{4}$, where the entropy remains at $1$. 
    Middle: Total single-domain wall sample proportion (agnostic to the spin orientation of the single domain wall). 
    Bottom: Total single-domain wall proportion (abbreviated as SDWP) which moved from the initial pinned position specified by the reverse anneal initial spin configuration (or has the opposite domain wall orientation to that initial domain wall). Each sub-plot shares the same x-axis scale. }
    \label{fig:domain_wall_entropy_2000_ms}
\end{figure*}

Quantum annealing is a type of analog quantum computation~\cite{Morita_2008, farhi2000quantumcomputationadiabaticevolution, Santoro_2002, Kadowaki_1998} based on the quantum adiabatic theorem, to efficiently find solutions of discrete combinatorial optimization problems. More generally, quantum annealers as an analog platform have emerged as a versatile programmable testbed for various aspects of condensed matter physics~\cite{king2024computationalsupremacyquantumsimulation, King_2021,King_2018,PhysRevB.104.L081107,lopez2023kagome,lopez2024quantum,sathe2025classical,PhysRevB.110.054432,pelofske2025magnetic}. We use here D-Wave Quantum Processing Units (QPU) based on superconducting flux qubits~\cite{Bunyk_2014, johnson2011quantum, dickson2013thermally}, governed by the Hamiltonian (in this case specifically applied to the 1-dimensional system)
\begin{align}
    {\mathcal H} =\frac{B(s)J} {2} \sum_{i}  \hat\sigma_z^{i} \hat\sigma_z^{i+1}
     - \frac{A(s)}{2}  \sum_i \hat\sigma_x^{i},
    \label{equation:DWave_QA_Hamiltonian}
\end{align}
where $J$ is a normalized dimensionless programmable parameter, $A(s)$ and $B(s)$ are machine-specific energy functions of the annealing parameter $s$, depicted in Fig.~\ref{fig:protocol_and_hardware_params}. It maps into that of Eq.~(\ref{equation:Transverse_Ising}) with $B J/2 \xrightarrow{} J$ and $A/2 \rightarrow \Gamma$. 

We study the effect of quantum fluctuations on an initial ``classical" configuration. After preparing the system in a one-domain-wall configuration in the ${\cal Z}$ basis, we increase $\Gamma$ and reduce $J$ by changing the annealing parameter $s$ as shown in Fig.~\ref{fig:protocol_and_hardware_params} (see also Supp. Mat.): a ramping up for 0.5 $\mu$s, followed by holding it for 2 (Fig~\ref{fig:domain_wall_entropy_2_ms}), 100 (Fig~\ref{fig:domain_wall_entropy_100_ms}), and 2000 (Fig~\ref{fig:domain_wall_entropy_2000_ms}) $\mu$s before bringing $\Gamma$ back to zero, and $J$ up to its programmed value (either $J=1$, or $J=0.001$). At these annealing times, the quantum processor is an open quantum system~\cite{king2022coherent, king2024computationalsupremacyquantumsimulation, king2023quantum}, with both thermalization effects and various sources of error that can affect the computation~\cite{PhysRevApplied.19.034053, Pelofske_2023_noise, Amin_2015, PhysRevApplied.8.064025, nelson2021singlequbitfidelityassessmentquantum, buffoni2020thermodynamics}. The D-Wave devices used in this study have coupler hardware graphs known as Pegasus~\cite{boothby2020nextgenerationtopologydwavequantum, dattani2019pegasussecondconnectivitygraph} and Zephyr~\cite{zephyr}. In order to map the 1D rings onto the physical hardware graph, we use a direct qubit to spin mapping. In particular an isomorphism to the native hardware graph is found using the subgraph isomorphism finder tool called Glasgow~\cite{mccreesh2020glasgow}. For all \texttt{Advantage\_system4.1} simulations, a $4905$ spin embedded ring is used, for all \texttt{Advantage\_system6.4} simulations a $4885$ spin ring is used, and for all \texttt{Advantage2\_system1.3} simulations a $4059$ spin ring is used.

The analog quantum simulation protocol we use is ``reverse quantum annealing'', first introduced under the name ``Sombrero Adiabatic Quantum Computation''~\cite{perdomo2010studyheuristicguessesadiabatic}, and studied extensively as a local search heuristic for improving on sub-optimal solutions to discrete combinatorial optimization problems~\cite{golden2021reverse, PhysRevA.101.022331, PhysRevA.98.022314, Pelofske_2023_reverse, Pelofske_2020_reverse, PhysRevA.102.062606, unraveling_RA, PhysRevA.104.012604}, as well as a Monte-Carlo-like simulator for equilibration simulations of magnetic systems~\cite{King_2021, PRXQuantum.2.030317, king2021scaling, PRXQuantum.1.020320}. Its performance is often assessed in the context of discrete combinatorial optimization, with annealing dynamics and performance most commonly measured by the Hamming distance~\cite{PhysRevResearch.5.013224, perdomo2010studyheuristicguessesadiabatic, PhysRevA.102.062606, unraveling_RA, PhysRevA.104.012604}, which is not well suited for measuring configuration memory in general for magnetic systems. We therefore propose Shannon entropy of the domain-wall distribution as a natural metric, as it captures domain-wall diffusion under increasing transverse field.

We utilize reverse quantum annealing with ``symmetric'' anneal schedules meaning that the pause of $s$ (see Eq.\eqref{equation:DWave_QA_Hamiltonian}) occurs in the middle of the anneal, and the quenches before and after the pause have the same duration. Example anneal schedules used are illustrated as anneal time vs $s$ plots in Figure~\ref{fig:protocol_and_hardware_params}-b. The simulation proceeds by re-initializing the same classical spin configuration for each anneal-readout cycle (this is done by specifying \texttt{reinitialize\_state} as True). We vary the pause anneal fraction from $s=0$ to $s=1$ in steps of $0.001$, meaning there are $1001$ different anneal pauses that are used. The initial state used in the reverse quantum annealing simulations specifies the location of a domain wall at an arbitrarily chosen edge in the 1D spin system, and in particular it is a spin up domain wall ($\uparrow \uparrow$). The other key parameter that we vary is the energy scale of the programmed $J$ value, which we set to $1.0$ and $0.001$ (this is accomplished by setting \texttt{auto\_scale} to False). $J=1.0$ is in normalized hardware units, representing the strongest coupling that can be programmed, at least for antiferromagnetic coupling, on the hardware. $J=0.001$ is on another extreme -- this is a very weak antiferromagnetic coupling which is likely very near to the precision limit on the analog hardware~\cite{pelofske2024simulatingheavyhextransversefield}.

For each value of $\Gamma/J$, $\tau$, and D-Wave QPU, we perform $8,000$ anneal-readout cycles, with the goal of mitigating finite sampling effects. From these $8,000$ samples we compute the probability $p_e$ of finding a domain wall on the edge $e$ between two qubits on the ring, normalized such that $\sum_e p_e=1$. (We do not differentiate between the domain wall orientation---spin up $\uparrow \uparrow$ or spin down $\downarrow \downarrow$.)
The Shannon \emph{Domain Wall Entropy} of $p_e$ is 
\be
h[p] = - \sum_{e} p_e \log_N (p_e),
\ee
where $N$ is the number of edges $e$ (and also the number of spins). 

The Shannon entropy quantifies the information gained on the outcome of the exposure to quantum fluctuations. For example, at sufficiently low $\Gamma/J$, the domain wall should remain stationary, no new domain walls are created, and the entropy is minimal, or $h = 0$. Conversely, for $\Gamma/J$ sufficiently large, quantum fluctuations produce domain walls. For enough domain walls production, any memory of the original one is lost leading to a uniform outcome distribution $p_e = 1/N$: an experiment yields maximal information gain, corresponding to $h = 1$. 
In an intermediate regime, quantum fluctuations merely displace the domain wall, resulting in a peaked distribution $p_e$ centered around the original position, henceforth denoted as $e=0$. This corresponds to a partial information gain, $0 < h < 1$. Note that this entropy measure does not take into account the \emph{proportion} of domain walls sampled out of all samples that were measured, but rather purely the \emph{distribution} of domain walls.

\begin{figure*}[ht!]
    \centering
    \includegraphics[width=0.49\linewidth]{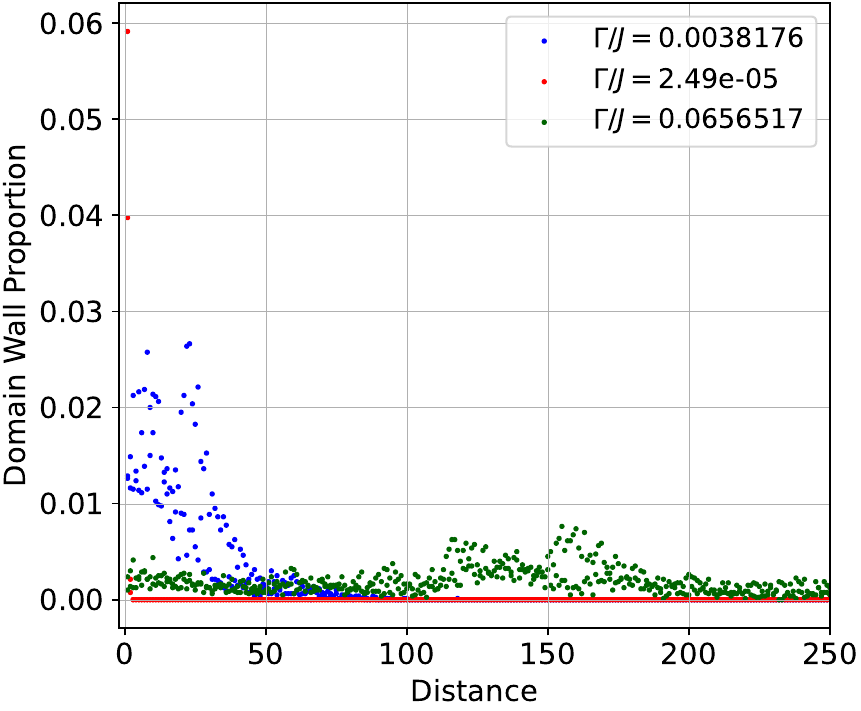}
    \includegraphics[width=0.49\linewidth]{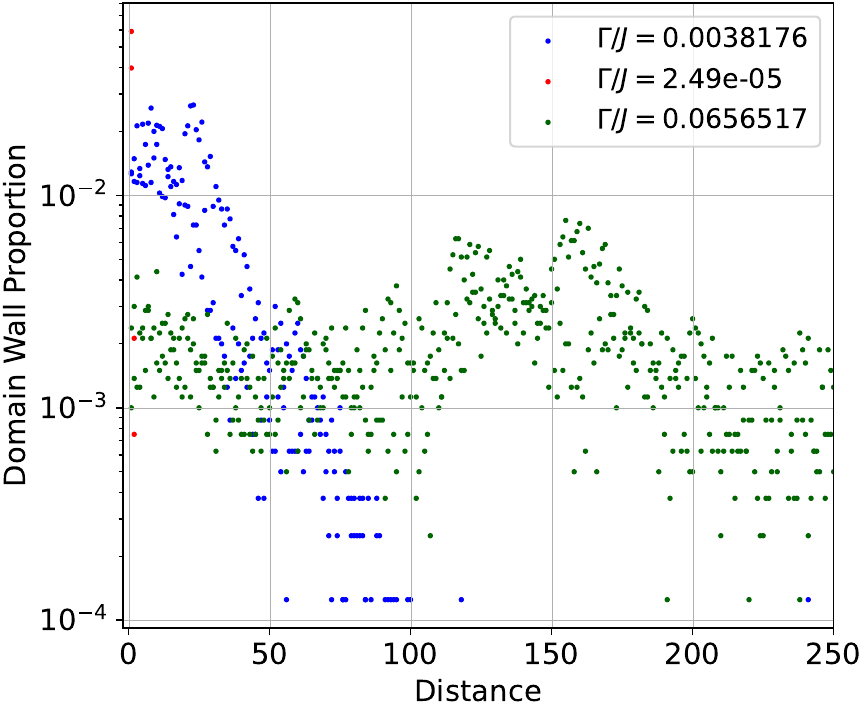}
    \includegraphics[width=0.49\linewidth]{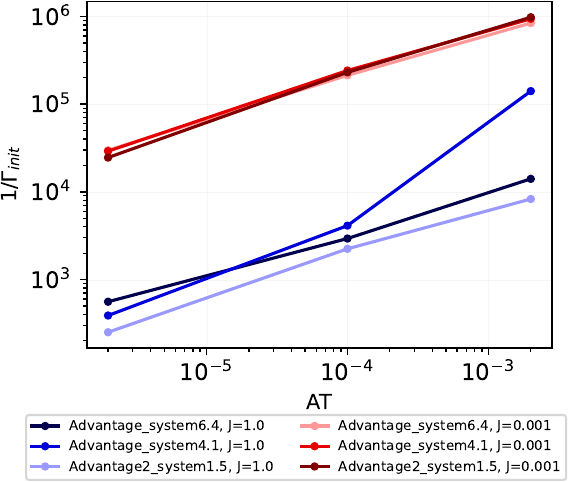}
    \includegraphics[width=0.49\linewidth]{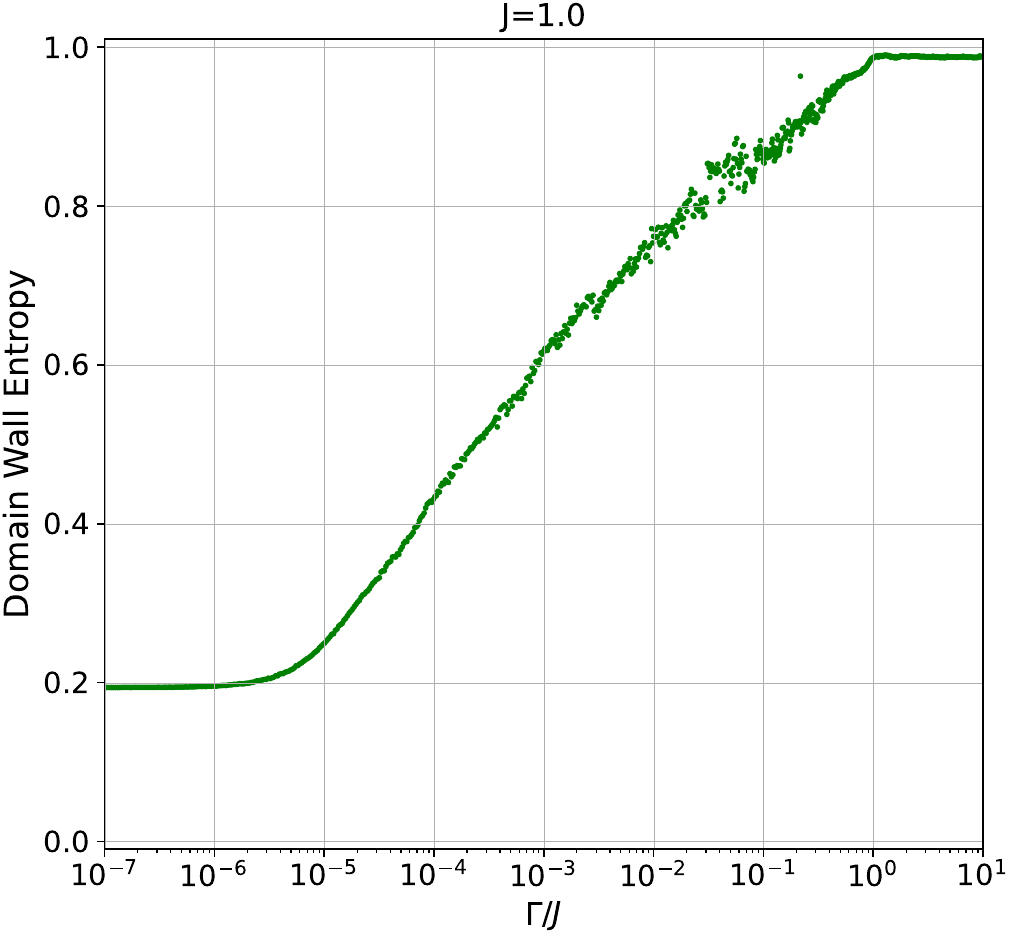}
    \caption{\textbf{Top row:} Spatial domain wall density distribution from three points on the \texttt{Advantage2\_system1.3} results from $J=1.0$, AT of $2000 \mu$ s: at the peak of non-initial domain wall densities, and before the peak and after the peak. Log y-axis scale version of the plot on the top right. The x-axis distance is artificially restricted to $250$ spins away from the initial pinned domain wall. 
    \textbf{Lower left:} Initial entropy change (by $\approx 0.05$ entropy change from zero), in terms of $1/\Gamma$, as a function of annealing time in seconds. Transverse field ($\Gamma_\text{init}$) in units of GHz. The best fit linear functions to this data, in log-log scale, have slopes of 0.465, 0.841, 0.509, 0.484, 0.505, 0.535.
    \textbf{Lower right:} Domain wall entropy results from \texttt{Advantage2\_system1.3}, $J=1$ and $2000$ $\mu$s, when two of the qubits in the ring, unexpectedly, began to malfunction during data collection, resulting in a domain wall entropy of $\approx 0.2$ when $\Gamma/J$ was very small (where we would expect the entropy to go to zero). Similarly to the prior experimental figures, the x-axis scale is cropped to include only the data in the region where the entropy changes. }
    \label{fig:domain_wall_location_and_initial_entropy_change_scaling_Fig5}
\end{figure*}

Figure~\ref{fig:domain_wall_entropy_2_ms}, \ref{fig:domain_wall_entropy_100_ms} and~\ref{fig:domain_wall_entropy_2000_ms}, show results for an exposure time of $2~\mu$s,  $100~\mu$s, and $2000~\mu$s, respectively, on three different D-Wave QPU's, summarized in Table~\ref{table:hardware_summary} in S.M.~\ref{section:appendix_DWave_hardware_details}. We plot the magnetic domain wall Shannon entropy, the probability of observing a single domain wall, and the probability that this domain wall has shifted from its initial location, all vs. $\Gamma/J$.  

In the very weakly interacting regime ($J = 0.001$), information reflects only the single-qubit dynamics of the system. The Shannon entropy displays a smooth sigmoidal dependence at all exposure times, providing a useful benchmark of quantum annealer behavior, fairly independent from the collective dynamics. Because creating a domain-wall pair incurs almost no energy cost, the sigmoid is narrower than in strongly coupled cases, as seen in the left columns of the three figures. This is consistent with the probability of finding a single domain wall, which at all times closely follows $1 - h$. By contrast, the probability of observing a (single) displaced domain wall remains negligible ($\sim 10^{-4}$) at all exposures. Finally, as expected, the sigmoidal shifts toward lower $\Gamma/J$ values with increasing exposure time. We fit the entropy transition to sigmoid curves in order to determine the domain wall entropy inflection point, more details are given in S.M.~\ref{section:appendix_methods_sigmoid_curve_fitting}. 

In this regime, \texttt{Advantage2\_system1.3} shows a considerably different behavior than the other two devices,
\texttt{Advantage\_system6.4} and \texttt{Advantage\_system4.1}, whose sigmoid curves are  almost indistinguishable. The difference might be due to a higher internal machine temperature, which varies slightly over time and across the different devices: \texttt{Advantage2\_system1.3} is at approximately $20$ milliKelvin, \texttt{Advantage\_system6.4} is at approximately $16$ milliKelvin, and \texttt{Advantage\_system4.1} is at approximately $15$ milliKelvin.

Relevant to us is the Window of Partial Memory (WPM), which describes how memory is erased, and is defined by $0 < h < 1$. Across all exposure times, the WPM is broader in the strongly coupled case than in the weakly coupled regime, and it grows in size with increasing exposure: spanning 4 orders of magnitude at $\tau = 2~\mu$s, 5 at $\tau = 100~\mu$s, and 6 at $\tau = 2000~\mu$s.

In the strongly coupled case, the WPM begins once $\Gamma$ is sufficient to move the initial domain wall. Domain wall motion accounts for most of the WPM span, as seen in the domain-wall statistics: in practice, most of the WPM corresponds to the regime of high probability for single domain-wall motion (lower and middle columns of Figs.~\ref{fig:domain_wall_entropy_2_ms}, \ref{fig:domain_wall_entropy_100_ms}, and \ref{fig:domain_wall_entropy_2000_ms}), with the maximum motion probability closely tracking $h=1/2$ for times $\tau=2$ and $100 \mu$s. Only when $\Gamma/J$ is large enough to create many new domain-wall pairs is the original memory lost and $h$ saturates to its $h=1$ maximum.

Figure~\ref{fig:domain_wall_location_and_initial_entropy_change_scaling_Fig5}, top-row, shows three representative $p_e$ distributions for \texttt{Advantage2\_system1.3}, strong $J=1$ coupling, at $\tau = 100~\mu$s, corresponding to the beginning, middle, and end of the WPM, illustrating the regime of domain-wall almost stationary (red), domain wall motion (blue) and domain-wall proliferation (green). 

Domain-wall motion arises from only flipping the adjacent spins without energy cost, and thus it does not depend on the value of $J$. However, the onset field $\Gamma_{\text{init}}$, defined arbitrarily as the value where $h = 0.05$, is orders of magnitude lower for the almost uncoupled case, because in that case thousands of qubits can flip freely, not merely the two impinging on the domain wall. Figure~\ref{fig:domain_wall_location_and_initial_entropy_change_scaling_Fig5} plots $1/ \Gamma_{\text{init}}$ versus $\tau$ for the three different D-Wave machines at both $J = 0.001$ and $J=1.0$, in log-log scale. While we are mindful of only having three data points, we can say that, with the exception of \texttt{Advantage\_system4.1} at very long times, we see an algebraic scaling consistent with $\Gamma_{\text{init}}\sim 1/\sqrt{\tau}$.

In the strongly coupled case, the entropy curves deviate from simple sigmoidal behavior. Here again \texttt{Advantage2\_system1.3} deviates even more strongly from the other quantum annealers, potentially suggesting that the increased thermal noise in the weakly coupled regime averages out local bias effects in the processor. Moreover, the strongly coupled regime might exhibit coupler leakage, a type of error source on the analog hardware. 

Interestingly, the case of a very long exposure time of $\tau=2000 \mu$s, shown in Figure~\ref{fig:domain_wall_entropy_2000_ms} leads to remarkably different behaviors. The entropy curves lose their smoothness and in fact are not even monotonic, with profiles differing across all QPU's. This suggests an accumulation and propagation of hardware errors and of coupling to the environment.

Finally, our methodology allows us to detect hardware problems. In Fig~\ref{fig:domain_wall_location_and_initial_entropy_change_scaling_Fig5} we show the Shannon entropy for \texttt{Advantage2\_system1.3} where two qubits on the hardware graph, both of which were part of the embedded 1D ring, unexpectedly malfunctioned during data collection (and were subsequently removed from the available working hardware graph). Because of that, at low $\Gamma/J$ the Shannon entropy never reaches zero but saturates at a fixed value $h=0.2$. In all other simulations run on this same QPU, we used a different minor-embedding, or the machine prior to the malfunction. Importantly, this shows the sensitivity of domain wall entropy to hardware error sources.

We have demonstrated a protocol that can be run on D-Wave quantum annealers which probes the ``effective memory'' of the transverse field Hamiltonian, in the form of the entropic distribution of pinned magnetic domain walls in an antiferromagnetic 1D chain.

In future work we aim to examine the sampling entropy of other types of frustrated topological features, such as pinned 2D domain walls induced by geometric frustration, using this same reverse quantum annealing simulation and subsequent magnetic domain wall entropy measurement on D-Wave quantum annealers. 

In conclusion, we have introduced the Shannon entropy of domain-wall distributions as a natural measure of memory retention under quantum fluctuations in the transverse Ising model, providing a device-agnostic benchmark for reverse annealing and a diagnostic of non-equilibrium dynamics. Our approach serves as a sensitive probe of hardware calibration and noise  furnishes a controlled testbed for comparing hardware behavior with theoretical predictions of quantum phase transitions, and informs the design of noise-resilient quantum memories by identifying regimes of partial or complete memory loss.
More broadly, controlled memory degradation under tunable fluctuations connects to key areas of quantum theory, including decoherence, non-Markovianity, fluctuation theorems, and the use of noise as a computational resource. By tracking memory decay in large spin systems, the method also links to questions of thermalization, scrambling, and many-body localization.

\section*{Acknowledgments}
\label{sec:acknowledgments}
This work was supported by the U.S. Department of Energy through the Los Alamos National Laboratory. Los Alamos National Laboratory is operated by Triad National Security, LLC, for the National Nuclear Security Administration of U.S. Department of Energy (Contract No. 89233218CNA000001). The research presented in this article was supported by the Laboratory Directed Research and Development program of Los Alamos National Laboratory under project number 20240032DR. This research used resources provided by the Los Alamos National Laboratory Institutional Computing Program. LA-UR-25-28344.

\renewcommand{\appendixname}{Supplementary Material}

\appendix

\section{D-Wave Quantum Annealer Hardware Details}
\label{section:appendix_DWave_hardware_details}

\begin{table*}[th!]
    \begin{center}
        \begin{tabular}{|l||l|l|l|p{3.3cm}|}
            \hline
            D-Wave QPU Chip & Graph name & Qubits & Couplers & Largest Odd-Spin Ring Embedding Found \\
            \hline
            \hline
            \texttt{Advantage\_system4.1} & Pegasus $P_{16}$ & 5627 & 40279 & 4905 \\
            \hline
            \texttt{Advantage\_system6.4} & Pegasus $P_{16}$ & 5612 & 40088 & 4885 \\
            \hline
            \texttt{Advantage2\_system1.3} & Zephyr $Z_{12}$ & 4597 & 41870 & 4059  \\
            \hline
        \end{tabular}
    \end{center}
    \caption{Summary of the three D-Wave QPU's used in this study, along with the largest odd-spin 1D systems that we found native graph isomorphism embeddings onto the QPU hardware graphs, using a reasonable amount of (classical embedding) compute time. }
    \label{table:hardware_summary}
\end{table*}

\begin{figure}[ht!]
    \centering
    \includegraphics[width=1.0\linewidth]{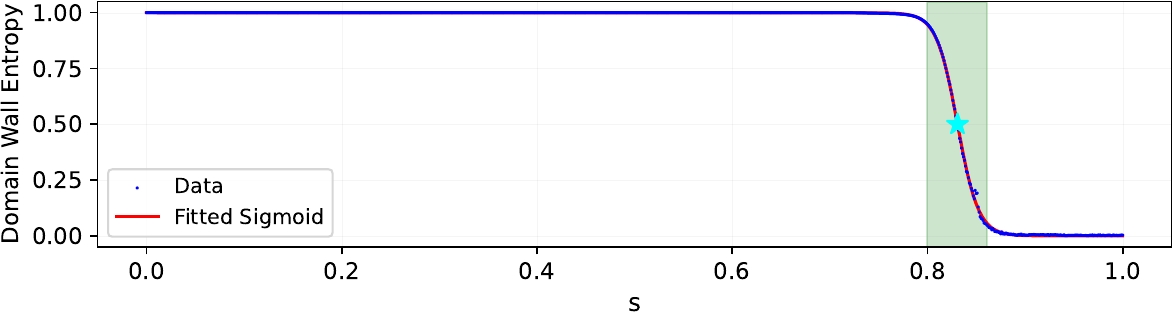}
    \includegraphics[width=1.0\linewidth]{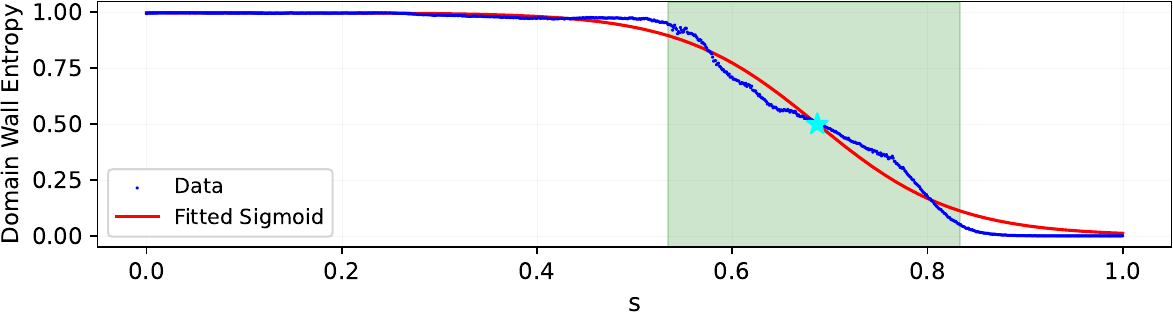}
    \caption{Sigmoid curve fits to the experimental D-Wave QPU data, in terms of the hardware-normalized anneal-fraction quantity $s$ (which is equivalent to a particular $\Gamma/J$ quantity). Cyan star denotes the inflection point of the fitted sigmoid curve. Blue is experimental data, and red is the continuous sigmoid curve fit. The shaded vertical green region shows the transition region, found by the region where the domain wall Shannon entropy is above $0.05$ and less than $0.95$. Top sub-plot is from \texttt{Advantage\_system4.1} at $J=0.001$, bottom sub-plot is from \texttt{Advantage2\_system1.3} at $J=1.0$.  }
    \label{fig:sigmoid_fits}
\end{figure}

\begin{figure*}[ht!]
    \centering
    \includegraphics[width=0.999\linewidth]{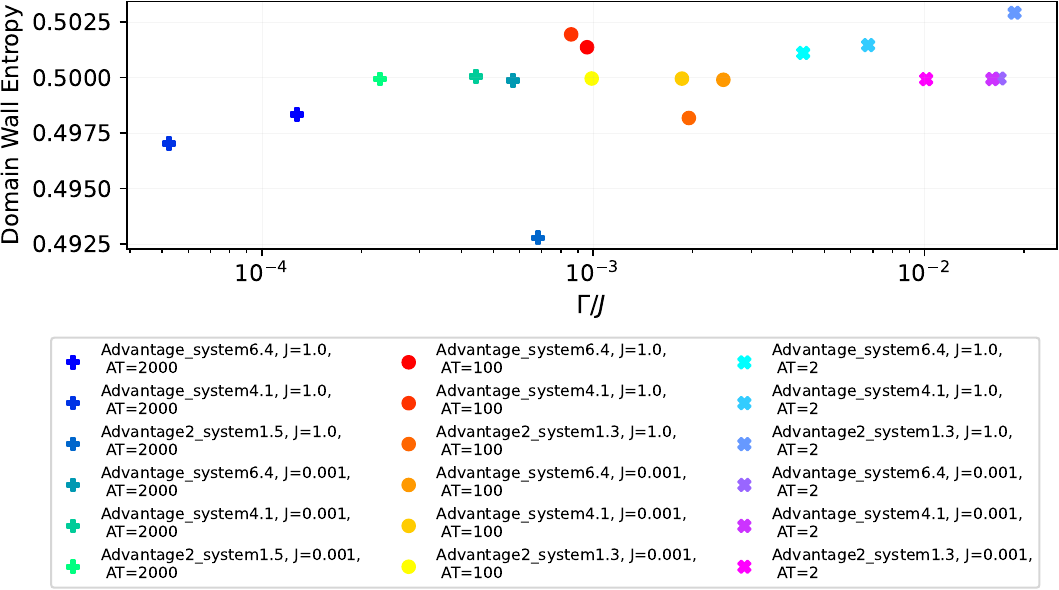}
    \caption{Domain wall entropy inflection points using on the best fitted sigmoid curve fit. There is a consistent shift in the inflection point to larger $\Gamma/J$ when the annealing time is shorter. }
    \label{fig:inflection_points}
\end{figure*}

The three D-Wave QPU's used are summarized in Table~\ref{table:hardware_summary}, along with the 1D spin system embedding, each with periodic boundary conditions, used in all D-Wave simulations shown in this study. Note that the QPU hardware graphs are defined as logical and tile-able structures, but there can be hardware defects that make the number of qubits and couplers present in the hardware graph fewer than the logical graph description.

As illustrated in Figure~\ref{fig:protocol_and_hardware_params}, we use initial and final ramps of duration $0.5$ microseconds -- ideally these ramps would be as short as possible, but there is a maximum slope hardware limitation that means for reading out from $s=0$ would require a quench to $s=1$ that could only be as fast as $0.5$ microseconds. And we use a quench duration of $0.5$ microseconds to be consistent for all simulations.

The $8,000$ samples per device setting are measured using $8$ separate device calls comprised each of $1,000$ anneal-readout cycles, with the exception of $2000$ $\mu$s anneals where $32$ device-calls comprised of $250$ anneal-readout cycles each were used in order to fit within the QPU-time-per-job constraints of the device. All other D-Wave QPU hardware settings not specified were left at their default values. 

The largest embedding reported in Table~\ref{table:hardware_summary} is the system we use for each D-Wave QPU, and is also the largest (odd) 1D spin system embedding that we found using a reasonable amount of CPU compute time. These embeddings are somewhat randomly structured in terms of layout on the hardware graph, and they make use a of a significant portion of the available qubits and couplers on the hardware graph. Note that during the sequence of experiments, \texttt{Advantage\_system1.3} had a hardware graph change, resulting in $4596$ available qubits and $41,851$ couplers, resulting in a different hardware embedding needing to be computed (also with exactly the same $4059$ spins, but the slightly different hardware was used). This change that occurred was the malfunctioning qubits that caused the entropic measure anomaly shown in Figure~\ref{fig:domain_wall_location_and_initial_entropy_change_scaling_Fig5}. Therefore, although the hardware changed, as did the exact chip id to reflect the hardware change, for consistency we refer to the processor as \texttt{Advantage\_system1.3} throughout the text.

\section{Sigmoid Curve Fitting and Domain Wall Entropy Inflection Points}
\label{section:appendix_methods_sigmoid_curve_fitting}

The quantum annealing simulation data exhibits a transition at particular values of $\Gamma/J$, and consequently the data resembles sigmoid curves. Therefore, a natural analysis of the data is to fit the measured domain wall entropy to best-fit sigmoid curves, in particular for the purpose of extracting the inflection point of the entropy transition from the sampled data found on hardware. The sigmoid model function we use is $y=\frac{L}{1 + \exp(-k \cdot (x-x_0))}$. This curve fitting is accomplished using a least-squares curve fitting using scipy~\cite{2020SciPy-NMeth}, and initial sigmoid parameters of $L=1.0, x_0=0.8, k=-40$. Figure~\ref{fig:sigmoid_fits} illustrates fitted sigmoid curves to a subset of the experimental data, showing how we can determine the inflection point of the entropic transition.

\begin{figure*}[ht!]
    \centering
    \includegraphics[width=0.999\linewidth]{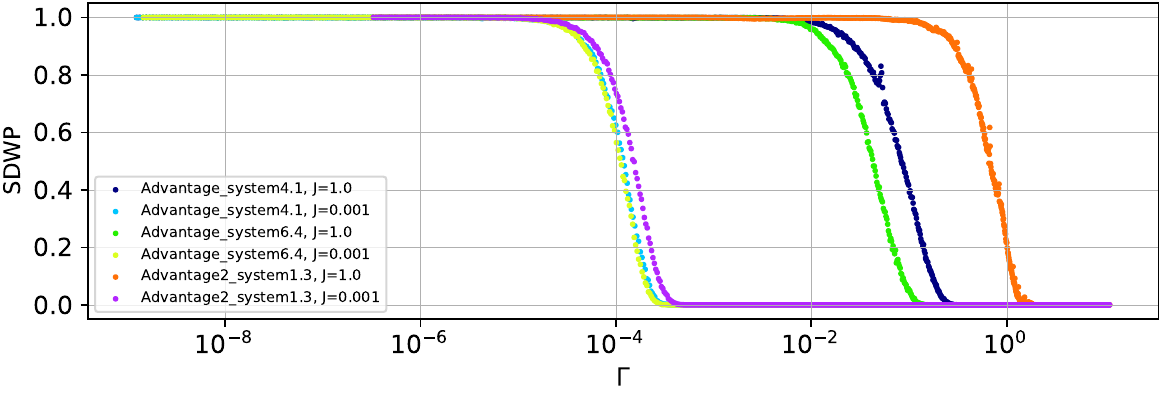}
    \includegraphics[width=0.999\linewidth]{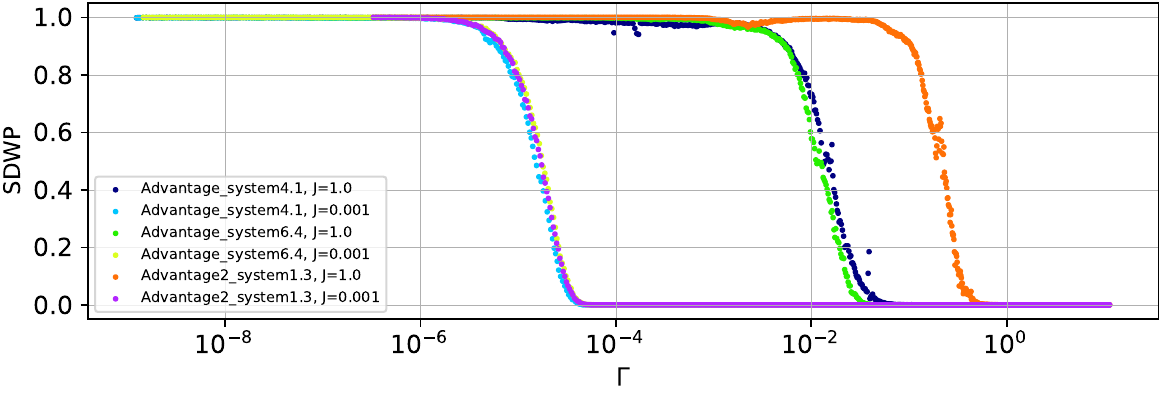}
    \includegraphics[width=0.999\linewidth]{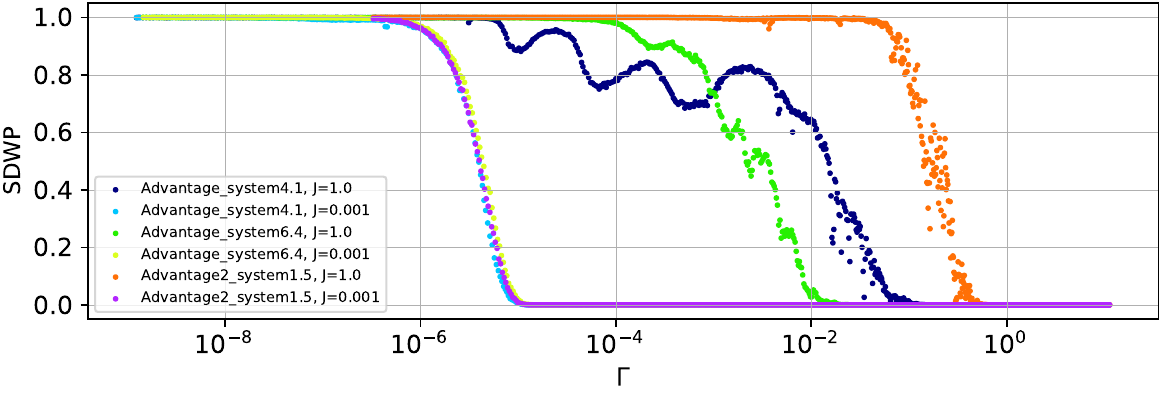}
    \caption{Single domain wall proportion, abbreviated as SDWP, (regardless of the spin of the domain wall, or the location wall) as a function of $\Gamma$, each sub-plot corresponds to a different annealing time; $2$ $\mu$s simulation time (top), $100$ $\mu$s (middle), and $2000 \mu$s (bottom). These plots represent the same single domain wall proportions in Fig.~\ref{fig:domain_wall_entropy_2_ms}, \ref{fig:domain_wall_entropy_100_ms}, \ref{fig:domain_wall_entropy_2000_ms}, but as a function of only the transverse field $A(s)$.  }
    \label{fig:single_domain_wall_proportion_function_of_Gamma}
\end{figure*}

\begin{figure*}[ht!]
    \centering
    \includegraphics[width=0.999\linewidth]{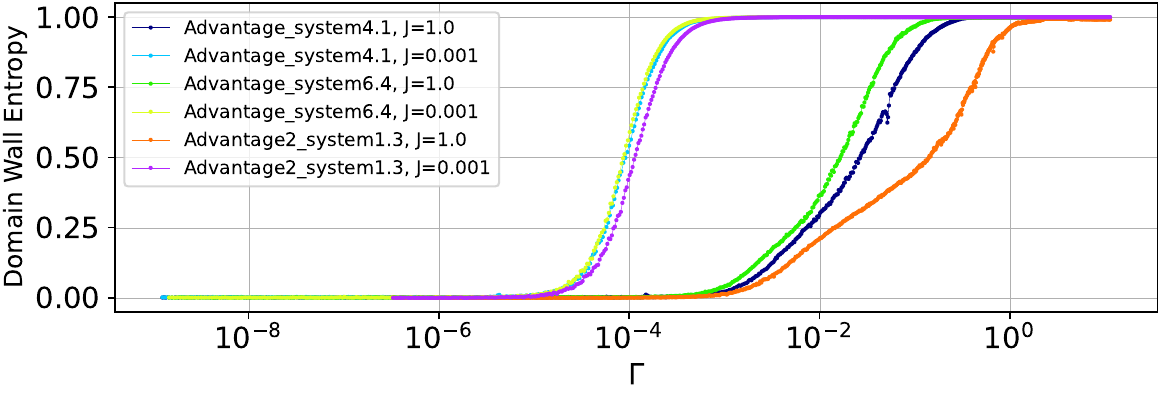}
    \includegraphics[width=0.999\linewidth]{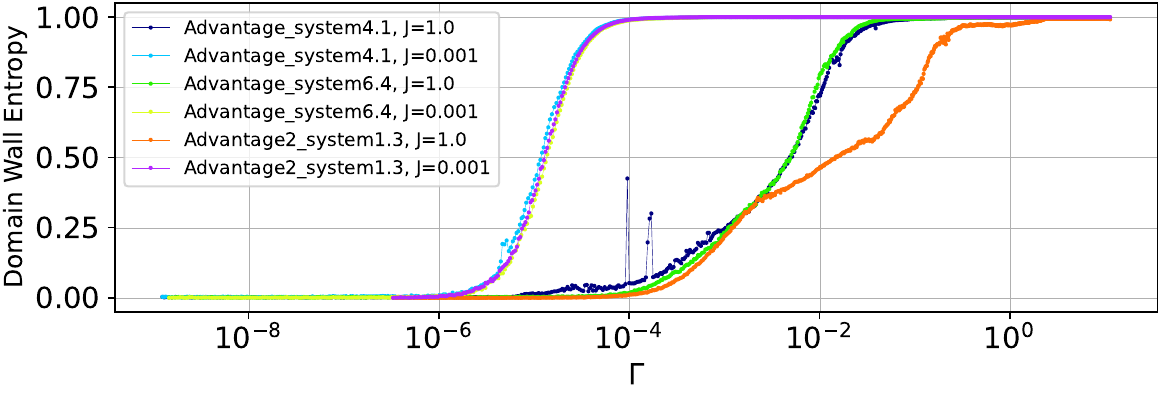}
    \includegraphics[width=0.999\linewidth]{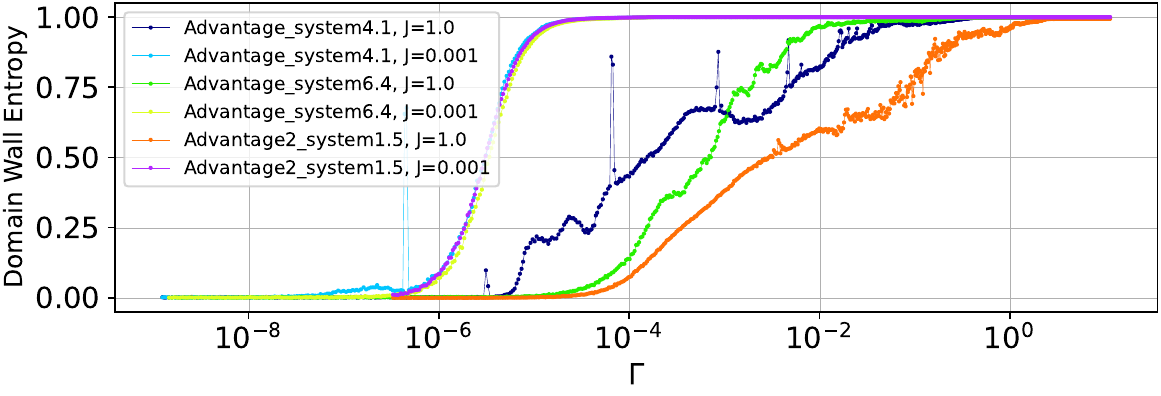}
    \caption{Domain wall entropy as a function of $\Gamma$, for a simulation time of $2 \mu$s (top), $100 
    \mu$s (middle), and $2000 
    \mu$s (bottom).  }
    \label{fig:domain_wall_entropy_function_of_Gamma}
\end{figure*}

Figure~\ref{fig:inflection_points} plots the inflection points across all experimental D-Wave QPU data. We see that the entropy inflection point, as expected, consistently occurs very close to $0.5$. Importantly, some of the data does not have a single inflection point, nonetheless this best-fit sigmoid curve fitting allows us to extract the global domain wall inflection point.

\section{Domain Wall Entropy and Proportion as a Function of Transverse Field}
\label{section:appendix_function_of_transverse_field_strengths}

Figure~\ref{fig:single_domain_wall_proportion_function_of_Gamma} presents the measured single domain wall proportion as a function only of $\Gamma$, instead of $\Gamma/J$. Similarly, Figure~\ref{fig:domain_wall_entropy_function_of_Gamma} shows the domain wall entropy as a function of $\Gamma$. The general trends that are observed are consistent with the $\Gamma/J$ scaling figures (Figs~\ref{fig:domain_wall_entropy_2_ms}, \ref{fig:domain_wall_entropy_100_ms}, \ref{fig:domain_wall_entropy_2000_ms}), but specifically highlight dynamics strictly as a function of the transverse field.

\begin{figure*}[ht!]
    \centering
    \includegraphics[width=0.999\linewidth]{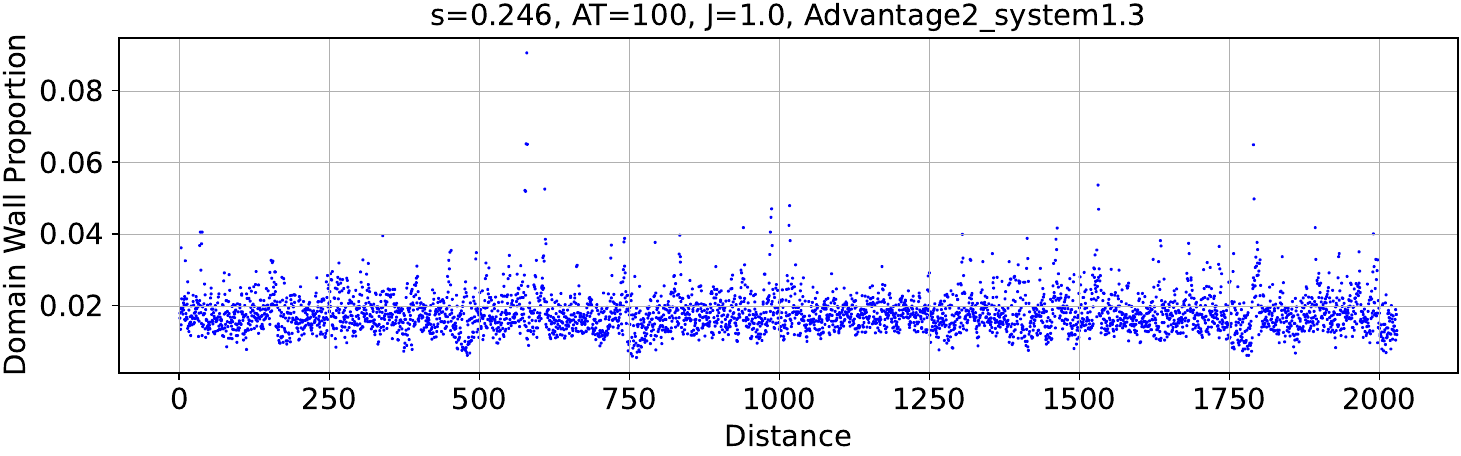}
    \includegraphics[width=0.999\linewidth]{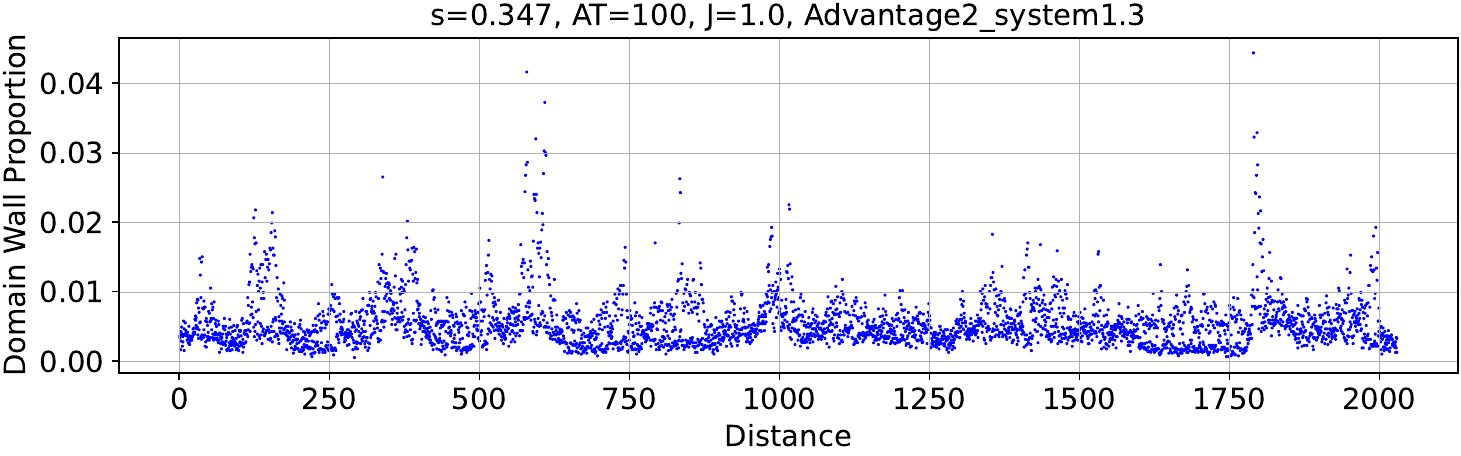}
    \includegraphics[width=0.999\linewidth]{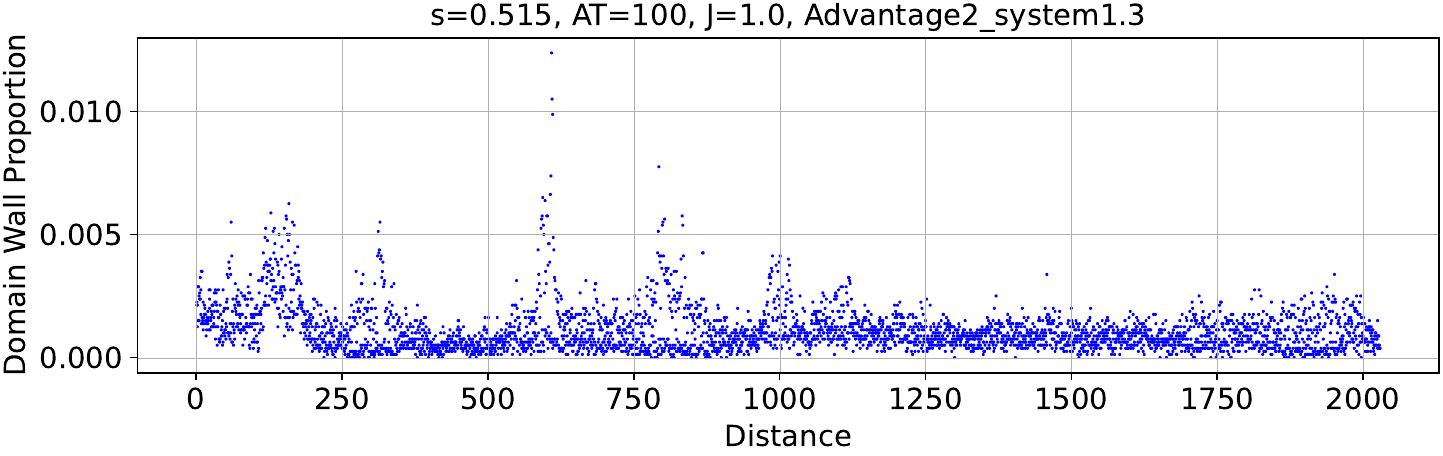}
    \includegraphics[width=0.999\linewidth]{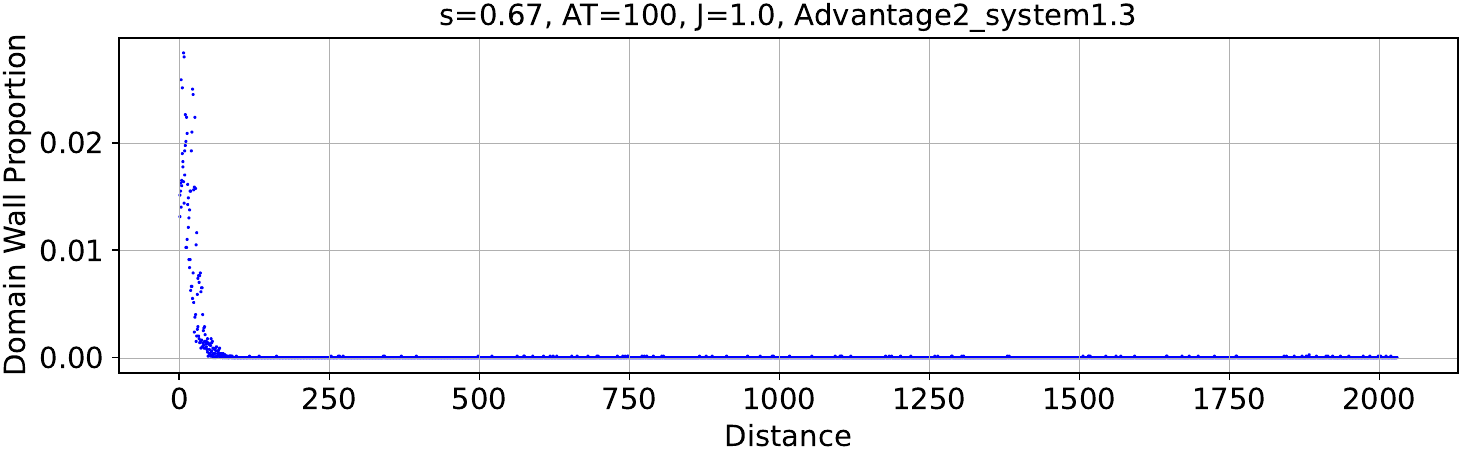}
    \caption{Domain wall density as a function of the distance from the initial reverse quantum annealing specified pinned domain wall. Here, orientation of the domain wall is not considered, only the position relative to the initial encoded domain wall. Each of these four sub-plots shows the domain wall density at four representative $s$ pause values (different ratios of $\Gamma/J$).  }
    \label{fig:spatial_domain_wall_density}
\end{figure*}

\section{Domain Wall Spatial Density Plots}
\label{section:appendix_domain_wall_density_spatial_distribution}

Figure~\ref{fig:spatial_domain_wall_density} shows four example measured domain wall densities from \texttt{Advantage2\_system1.3} (therefore, a ring with $4059$ spins) at $J=1$ coupling and $100 \mu$s simulation time. Interestingly, we see in the intermediate domain wall entropy regime many domain wall density fluctuations, whereas at larger $s$ values (smaller $\Gamma/J$), we see effectively no domain walls far from the initial pinned domain wall. We expect that most of the domain wall density peaks and fluctuations are due to analog noise on the processor, or other types of noise such as open quantum system effects. Additional analysis of these types of domain wall density distributions will be a subject of future work. In Figure~\ref{fig:spatial_domain_wall_density}, the domain wall density excludes domain walls (either spin up, or spin down) that are in the location in the ring where the reverse quantum annealing spin configuration specified the single domain wall -- this is to make the domain wall density clearer visually, since the density at that initial location peaks close to $1.0$ when $\Gamma/J$ is small.